\newif\ifpdf
\ifx\pdfoutput\undefined
   \pdffalse
\else
   \pdfoutput=1
   \pdftrue
\fi

\ifpdf
\documentclass[11pt,a4paper]{article}
\else
\documentclass[11pt,a4paper, dvips]{article}
\fi

\usepackage{latexsym}
\usepackage{amssymb}
\usepackage{amsmath}
\usepackage[authoryear]{natbib}

\ifpdf
    \usepackage[pdftex]{graphicx}
    \usepackage{color}
    \definecolor{myred}{rgb}{0.5,0,0}
    \definecolor{myblue}{rgb}{0,0,0.75}
    \definecolor{mygreen}{rgb}{0,0.5,0}
   \usepackage[pdftex,
               pdftitle={The single risk factor approach to capital charges in case of
                        correlated loss given default rates},
               pdfsubject={Regulatory capital requirements; loss given default},
               pdfkeywords={Capital charge; generalized inverse},
               pdfauthor={Dirk Tasche},
               pdfstartview=FitBH,
               breaklinks=true,
               colorlinks=true,
               citecolor=myred,
               linkcolor=myblue,
               urlcolor=mygreen]{hyperref}
\else
    \usepackage[dvips]{graphicx}
   \usepackage{hyperref}
\fi

\numberwithin{equation}{section}

\begin{document}
\title{The single risk factor approach to capital charges in case of correlated loss given default rates%
}

\author{%
Dirk Tasche\thanks{Deutsche Bundesbank, Postfach 10 06 02, 60006 Frankfurt am
Main, Germany\newline E-mail: tasche@ma.tum.de}\ %
\thanks{The opinions expressed in this note are those of the
author and do not necessarily reflect views of the Deutsche Bundesbank. The
author thanks Katja Pluto for helpful comments on an earlier version of
this note.} }

\date{February 17, 2004}
\maketitle

\begin{abstract}
A new methodology for incorporating LGD correlation effects into the Basel~II
risk weight functions is introduced. This methodology is based on modelling of
LGD and default event with a single loss variable. The resulting formulas for
capital charges are numerically compared to the current proposals by the Basel
Committee on Banking Supervision.
\end{abstract}

\section{Introduction}
\label{sec:intro}
Up to now (January 2004), the proposals of the Basel Committee on Banking
Supervision (BCBS) for new rules of regulatory capital requirements (Basel II)
are implicitly based on an assumption of independent loss given default (LGD)
rates and default events. Since the actual risk weight formulas were derived
by means of a transition to an asymptotic limit that eliminated diversifiable
-- and in particular independent -- components, LGDs enter the risk weight
formulas only by their expected values. However, over the past years evidence
of correlation effects between LGDs and default events became stronger
\citep[cf.][]{Frye00, Frye03}.

In the literature \citep{Pykhtin03}, an approach to incorporating LGD
correlation into the Basel II one-factor model is suggested which models LGDs
as \emph{potential losses} that are defined but not observed also in case of
the obligor being non-defaulted. Apart from this conceptual difficulty,
another problem of the potential losses approach follows from the need to
simultaneously estimate an additional -- compared to the Basel II model --
correlation parameter.

By modelling LGD dependence via the conceptually different \emph{single risk
factor} approach, this note suggests a way to avoid the problems which are
encountered with the model by \citeauthor{Pykhtin03}.
In particular, the single risk factor model does not require additional
correlation parameters but can be fed with LGD volatilities whose values can
be statistically estimated by the banks or prudentially fixed by the
supervisory authorities. Additionally, the model shares with
\citeauthor{Pykhtin03}'s model the property of
extending the Basel II model in a way that takes into account the case of
defaulted obligors (i.e.\ obligors whose probability of default equals 1).

This note is organized as follows: The single risk factor model of correlated
LGDs is derived in Section~\ref{sec:copula}. In particular,
Equation~\eqref{eq:approx} is ready for application as generic form of a
capital charge formula that recognizes LGD correlations. The question how to
calibrate the model is discussed in Section~\ref{sec:cali}. Numerical
illustrations of the concept are provided in Section~\ref{sec:num}.
Section~\ref{sec:concl} summarizes and concludes the note. In Appendix
\ref{sec:app_A}, we discuss in more detail the connection of the single risk
factor approach with the potential loss model.

\section{The single risk factor approach to regulatory LGD modelling}
\label{sec:copula}
The basic idea for the single risk factor approach is to describe with a
single loss variable the loss suffered with an obligor. This is in contrast to
the potential loss model \citep{Pykhtin03} where default event and amount of
loss are specified with separate random variables.

Consider a fixed obligor and denote by $L$ the loss as percentage of the total
exposure which results from granting credit to the obligor. Denote with $p$
the obligor's probability of default. Then $L$ will be zero with probability
$1-p$ and will take positive values with probability $p$. Formally, we
describe the probability distribution function $F_L$ of $L$ as
\begin{equation}\label{eq:dist}
  F_L(t) \ =\ \mathrm{P}[L\le t]\ =\ 1-p + p\,F_D(t),
\end{equation}
where $p = \mathrm{P}[L >0]$ is the obligor's probability of default and
$F_D(t) = \mathrm{P}[L\le t\,|\,L > 0] = p^{-1}\,\mathrm{P}[0 < L\le t]$ is
the distribution function of the observed losses. Denote with $\Phi$ the
standard normal distribution function. Then, in a Basel-II-like framework, $L$
can be represented as
\begin{equation}\label{eq:repr}
  L\ =\ F_L^\ast\bigl(\Phi(\sqrt{\rho}\,X+\sqrt{1-\rho}\,\xi)\bigr)
\end{equation}
with independent, standard normally distributed random variables $X$ and
$\xi$, \emph{asset correlation}\footnote{%
The variable $\sqrt{\rho}\,X+\sqrt{1-\rho}\,\xi$ is commonly interpreted as
minus the change in value of the obligor's assets within a fixed period of
time. If the loss expressed by the change is too large, the obligor defaults.
In order to express the dependence between different obligors, the change
variable is decomposed into a \emph{systematic} and an \emph{idiosyncratic}
part. The correlation $\rho$ of two change variables is called \emph{asset
correlation}. In technical terms, \eqref{eq:repr} may be interpreted as a
\emph{copula}-representation of the model \citep[cf.][]{Nel99}.%
}
 $\rho\in[0,1]$, and the
\emph{generalized inverse} or \emph{quantile} function $F_L^\ast$ of $F_L$. In
general, the quantile function $G^\ast$ of a distribution function $G$ is
defined as
\begin{equation}\label{eq:quantile}
  G^\ast(z)\ =\ \min\bigl\{t:\ G(t) \ge z\bigr\}.
\end{equation}
If $G$ is continuous and strictly increasing, then $G^\ast$ coincides with the
common inverse function of $G$, i.e.\ we have $G^\ast = G^{-1}$. Applied to
$F_L^\ast$, definition (\ref{eq:quantile}) yields the representation
\begin{equation}\label{eq:F.ast}
F_L^\ast(z)\ =\
  \begin{cases}
    0, & \text{if $z \le 1-p$}, \\
    F_D^\ast\bigl(\frac{z-1+p}{p}\bigr), & \text{if $z > 1-p$}
  \end{cases}
\end{equation}
for the quantile function of the obligor's loss distribution where $F_D^\ast$
again has to be determined according to \eqref{eq:quantile}. Together,
\eqref{eq:repr} and \eqref{eq:F.ast} imply
\begin{equation}\label{eq:product}
  L \ =\
  \begin{cases}
    0, & \text{if } \sqrt{\rho}\,X+\sqrt{1-\rho}\,\xi \le \Phi^{-1}(1-p), \\
    F_D^\ast\bigl(\frac{\Phi(\sqrt{\rho}\,X+\sqrt{1-\rho}\,\xi)-1+p}{p}\bigr), & \text{otherwise}.
  \end{cases}
\end{equation}
\eqref{eq:product} can be equivalently written as the product of the indicator
function of the default event $\sqrt{\rho}\,X+\sqrt{1-\rho}\,\xi >
\Phi^{-1}(1-p)$ and the factor
$F_D^\ast\bigl(\frac{\Phi(\sqrt{\rho}\,X+\sqrt{1-\rho}\,\xi)-1+p}{p}\bigr)$.
The second factor can -- similarly to
\citeauthor{Pykhtin03}'s model -- be interpreted as loss in case of default.

From representation \eqref{eq:dist} of the loss distribution  follows that in
the single risk factor model the expected loss may be calculated as the product
of the probability of default and the expected loss given default, i.e.
\begin{subequations}
\begin{equation}\label{eq:exloss}
  \mathrm{E}[L]\ =\ p\,\mathrm{E}[L\,|\,L>0]\ =\ p \int_0^\infty t\,F_D(d\,t).
\end{equation}
In order to be able to calculate capital charges in the Basel II sense, we
need in addition an evaluable representation of $\mathrm{E}[L\,|\,X=x]$
\citep[see, e.g.,][for an explanation]{Gordy01}. From \eqref{eq:product}
follows
\begin{equation}
  \mathrm{E}[L\,|\,X=x] \ = \ \int\limits_{\frac{\Phi^{-1}(1-p) - \sqrt{\rho}\,x}
  {\sqrt{1-\rho}}}^\infty \varphi(z)\,
  F_D^\ast\left(\frac{\Phi(\sqrt{\rho}\,x+\sqrt{1-\rho}\,z)-1+p}{p}\right)\,d
  z, \label{eq:cond}
\end{equation}
\end{subequations}
where $\varphi(z) = \bigl(\sqrt{2\,\pi}\,\bigr)^{-1}\, e^{- 1/2\, z^2}$
denotes the standard normal density. The case of an already defaulted obligor
($p=1$) is admitted in \eqref{eq:exloss} and \eqref{eq:cond}. The lower
integration limit in \eqref{eq:cond} then has to be taken as $-\infty$. Note
that the choice of $F_D$ as
\begin{subequations}
\begin{equation}\label{eq:constant}
F_D(t)\ =\
  \begin{cases}
    0, & \text{if $t < LGD$}, \\
    1, & \text{if $t \ge LGD$}.
  \end{cases}
\end{equation}
for some constant $LGD$ will bring us back to the Basel II framework of the
risk weight functions \citep[see][]{BC03} since then $F_D^\ast$ is simplified
to
\begin{equation}\label{eq:basel}
F_D^\ast(z)\ =\ LGD\qquad\text{for all}\ z \in (0,1).
\end{equation}
\end{subequations}
In Section~\ref{sec:cali}, we will show with an example how $F_D^\ast$ can be
chosen in order to incorporate correlation effects into the Basel risk weight
functions.

With recourse to Gauss quadrature the integral in \eqref{eq:cond} can be
efficiently approximated with only five addends \citep[cf.][]{Martin01}.
Observe that with the change of variable
\begin{equation}\label{eq:change}
  t \ = \ -\,1+2\,p^{-1}\,\bigl(\Phi(\sqrt{\rho}\,x+\sqrt{1-\rho}\,z)-1+p\bigr)
\end{equation}
the integral on the right-hand side of \eqref{eq:cond} can be written as
\begin{subequations}
\begin{align}\label{eq:gauss}
\mathrm{E}[L\,|\,X=x] & =  \frac{p}{2\,\sqrt{1-\rho}} \int_{-1}^1 H(t,x)\,d
t\\
\intertext{with} H(t,x) & =
\frac{\varphi\left(\frac{\Phi^{-1}(p\,(t+1)/2+1-p)-\sqrt{\rho}\,x}{\sqrt{1-\rho}}\right)}
{\varphi\bigl(\Phi^{-1}(p\,(t+1)/2)+1-p\bigr)}\,
F_D^\ast\bigl((t+1)/2\bigr).\label{eq:integrand}
\end{align}
\end{subequations}
Representation \eqref{eq:gauss}, \eqref{eq:integrand} is appropriate for a
Gauss-approximation with weights and sampling points derived from the Legendre
polynomials \citep[cf.][]{Stoer}. In particular, an approximation of
$\mathrm{E}[L\,|\,X=x]$ with five weights and sampling points is given by
\begin{subequations}
\begin{equation}\label{eq:approx}
  \mathrm{E}[L\,|\,X=x] \ \approx\ \frac{p}{2\,\sqrt{1-\rho}} \sum_{i=1}^5
  w_i\,H(t_i, x),
\end{equation}
where the numbers $w_i$ and $t_i$ are specified as
\begin{align}
  t_1 & = -0.9061798459, & w_1 & = 0.2369268851,\notag \\
  t_2 & = -0.5384693101, & w_2 & = 0.4786286705,\notag\\
  t_3 & = 0, & w_3 & = 128/225,\label{eq:weights}\\
  t_4 & = - t_2, & w_4 & = w_2,\notag\\
  t_5 & = - t_1, & w_5 & = w_1.\notag
\end{align}
\end{subequations}

\section{Calibrating the model}
\label{sec:cali}
According to the current Basel II philosophy\footnote{%
Charging capital only for the \emph{unexpected loss} as expressed by the
difference of a high confidence level quantile of the loss and the expected
loss \citep{BC04}.},
the capital charge\footnote{%
For the sake of clarity, in this note we do not mention \emph{exposures at
default} and \emph{maturity adjustments}. However, the expressions for the
capital charges only have to be multiplied with the corresponding factors from
the Basel II model in order to cover the full scope of an extended Basel II
model.%
} as a percentage of the total exposure for the percentage loss variable $L$
will be calculated as
\begin{equation}\label{eq:charge}
  \text{Charge}(L) \ = \ \mathrm{E}[L\,|\,X=\Phi^{-1}(\alpha)] -
  \mathrm{E}[L],
\end{equation}
where $\alpha$ denotes some high confidence level  (currently $\alpha =
0.999$). In order to evaluate \eqref{eq:charge}, according to
\eqref{eq:exloss} and \eqref{eq:cond}, we have to specify values for the
probability of default $p$ and the asset correlation $\rho$ as well as to
provide an appropriate functional form for the distribution function $F_D$ of
the realized losses.

The parameters $p$ and $\rho$ can be estimated (or in case of $\rho$ set by
the supervisors) as usual in the Basel II framework. In principle, $F_D$ can
be estimated parametrically or non-parametrically from a sample of realized
losses. We suggest to choose a Beta-distribution\footnote{%
The same calculations as those presented in Section~\ref{sec:num} were carried
out with normal and gamma distributions instead of the Beta-distribution. The
observed differences in the resulting capital charges were negligible (less
than 2.5\% of the Basel~II charges).} as a parametric representation of $F_D$
and to determine its shape parameters $a$ and $b$ via moment matching from
estimates $LGD$ of the expectation of $F_D$ and $VLGD$ of the variance of
$F_D$. The general density of a Beta-distribution is given by
\begin{equation}\label{eq:dens}
  \beta(a,b;x) \ =\ \frac{\Gamma(a+b)}{\Gamma(a)\,\Gamma(b)}\,x^{a-1}\,(1-x)^{b-1}, \ 0 < x < 1,
\end{equation}
where $\Gamma$ denotes the Gamma-function expanding the factorial
  function to the positive real numbers.
For given values of $LGD$ and $VLGD$ the parameters $a$ and $b$ then can be
calculated by
\begin{subequations}
\begin{align}\label{eq:a}
  a & = \frac{LGD}{VLGD}\,\bigl(LGD\,
  (1-LGD)-VLGD\bigr) \\
  \intertext{and}
  b & = \frac{1-LGD}{VLGD}\,\bigl(LGD\,
  (1-LGD)-VLGD\bigr).\label{eq:b}
\end{align}
\end{subequations}
For regulatory purposes, it might be adequate to prescribe conservative values
for $VLGD$ instead of allowing to insert statistical estimates into \eqref{eq:a} and
\eqref{eq:b}. This can be conveniently afforded by specifying $VLGD$ as a
fixed percentage $v$ of the maximally possible variance $LGD\,
  (1-LGD)$ of $F_D$. This leads to the representations
\begin{subequations}
\begin{align}\label{eq:av}
  a & = LGD\,\frac{1-v}v \\
  \intertext{and}
  b & = (1-LGD)\,\frac{1-v}v.\label{eq:bv}
\end{align}
\end{subequations}
In current credit portfolio models, $v = 0.25$ seems to be a common choice
that we will adopt for the numerical examples in Section~\ref{sec:num}.
However, other values of $v$ might turn out to be more appropriate. Moreover,
as with the asset correlation, $v$ could be modelled as a function of the
probability of default or of the expected loss given default.
\begin{table}[ht]
\begin{center}
\begin{tabular}{|c|c|c|c|}  \hline
PD & Basel II charge & Single risk factor  & Approx.\ single risk  \\
 & & charge & factor charge\\
\hline\hline
0.03 & 0.6 & 0.7 & 0.7 \\ \hline 0.1 & 1.4 & 1.7 & 1.7\\ \hline 0.25 & 2.8 &
3.2 & 3.2\\ \hline 0.5 & 4.2 & 4.9 & 4.8 \\ \hline 0.75 & 5.1 & 6.1 & 6.0
\\ \hline
1.0 & 5.9 & 7.0 & 6.8\\ \hline 2.0 & 7.7 & 9.3 & 9.1 \\ \hline 3.0 & 8.8 &
10.8 & 10.5 \\ \hline 5.0 & 10.6 & 13.2 & 12.8 \\ \hline 7.5 & 12.5 & 16.0 &
15.4
\\ \hline 10.0 & 14.1 & 18.4 & 17.7 \\ \hline 15.0 & 16.4 & 22.5 & 21.5
\\ \hline
20.0 & 17.8 & 25.5 & 24.2 \\ \hline 100.0 & 0.0 & 25.9 & 20.6\\ \hline
\end{tabular}
\end{center}
{\small\sl Table~1: Capital charges according to Basel~II and single
  risk factor
approach as functions of PD. Expected LGD fixed at 45\%. LGD variance fixed at
25\% of maximally possible variance. Maturity fixed at one year. Asset
correlations calculated according to the Basel II corporate curve. All figures
in \%.} \label{tb:1}
\end{table}

\section{Numerical examples}
\label{sec:num}
In order to illustrate the effect caused by replacing the current Basel~II
formulas for the capital charges with \eqref{eq:exloss} and \eqref{eq:cond},
we calculated capital charges with both approaches as well as with the
approximation formula \eqref{eq:approx} as functions of the probability of
default and of the expected loss given default respectively. As described in
Section~\ref{sec:cali} we chose a Beta-distribution for modelling the
distribution of the loss given default rates. The parameters of this
Beta-distribution were fixed by \eqref{eq:av} and \eqref{eq:bv} with $v=0.25$.
The correlations involved in \eqref{eq:cond} were set according to the rule
for corporate exposures \citep{BC03}, i.e.
\begin{equation}
  \label{eq:corr}
  \rho\ =\ \rho(p)\ =\  \frac{1-e^{-50\,p}}{1-e^{-50}}\,0.12 +
\bigg(1-\frac{1-e^{-50\,p}}{1-e^{-50}}\bigg) 0.24.
\end{equation}
The calculations
in case of the single risk factor approach were exact in the sense of being
carried out with a high-precision numerical integration routine.

Table~1 lists
the results in the case of varying PDs with fixed expected LGD. Table~2 gives
the results in the case of fixed PD with varying expected LGD. In particular,
the tabulated numbers show that the relative difference of the Basel charges
and the ``single risk factor'' charges increases in PD and decreases in
expected LGD. Differences diminish to zero when the expected LGD approaches
100\%. With the exception of the 100\% PD case, the quality of the
\eqref{eq:approx} approximation is satisfactory. Even in this case, the
resulting charge of 20\% is much more realistic than the 0\% Basel~II charge.
\begin{table}[ht]
\begin{center}
\begin{tabular}{|c|c|c|c|}  \hline
LGD & Basel II charge & Single risk factor  & Approx.\ single risk  \\
& & charge & factor charge \\ \hline\hline
5 & 0.7 & 1.2 & 1.0\\ \hline 10 & 1.3 & 2.1 & 1.9\\
\hline 15 & 2.0 & 2.9 & 2.7 \\ \hline 20 & 2.6 & 3.6 & 3.5 \\ \hline 25 & 3.3
& 4.4 & 4.2
\\ \hline 30 & 3.9 & 5.1 & 4.9 \\ \hline 35 & 4.6 & 5.7 & 5.5 \\ \hline 40 &
5.2 & 6.4 & 6.2 \\ \hline 45 & 5.9 & 7.0 & 6.8 \\ \hline 50 & 6.5 & 7.6
& 7.5 \\ \hline 55 & 7.2 & 8.3 & 8.1 \\ \hline 60 & 7.8 & 8.8 & 8.7 \\
\hline 65 & 8.5 & 9.4 & 9.3 \\ \hline 70 & 9.1 & 10.0 & 9.9 \\ \hline 75 & 9.8
& 10.6 & 10.4 \\ \hline 80 & 10.4 & 11.1 & 11.0 \\ \hline 85 & 11.1 & 11.6 &
11.5 \\ \hline 90 & 11.7 & 12.1 & 12.0 \\ \hline 95 & 12.4 & 12.6 & 12.5 \\
\hline 100 & 13.0 & 13.0 & 13.0 \\ \hline
\end{tabular}
\end{center}
{\small\sl Table~2: Capital charges according to Basel~II and single
  risk factor
approach
as functions of expected LGD. PD fixed at 1.0\%. LGD variance calculated as
25\% of maximally possible variance. Maturity fixed at one year. Asset
correlations calculated according to the Basel~II corporate curve. All figures
in \%.} \label{tb:2}
\end{table}

\section{Conclusions}
\label{sec:concl}
We have introduced a new methodology for incorporating LGD correlation effects
into the Basel II capital charge formulas. The new methodology is based on a
single risk factor modelling of LGD and default event. Compared to the current
capital charge formulas, the new formulas require the specification of one
additional parameter, namely the LGD volatility. This LGD volatility can be
estimated with standard methods or be prescribed by the supervisors in order
to ensure a conservative regulatory capital allocation. By numerical examples,
we have illustrated that the difference of the capital charges according to
Basel~II and according to the single risk factor approach increases with
growing PDs and decreases with growing expected LGDs. The integral formulas
for the capital charge that are involved with the single risk factor model can
be efficiently approximated by a weighted sum -- as specified in Equation
\eqref{eq:approx} -- with five addends.


\appendix
\section{The potential loss model and its connection with the single risk factor model}
\label{sec:app_A}
We present an interpretation of the potential loss model by
\cite{Pykhtin03} that makes the model comparable to the single risk factor
model as described by \eqref{eq:product}. For the potential loss model, one
assumes that the loss experienced with a fixed obligor is driven by two
risk factor random variables: $U$ for toggling the default event and $V$ for
settling the amount of loss. Consequently, the loss variable $L$ is defined by
\begin{equation}\label{eq:A_L}
  L\ =\
  \begin{cases}
    G(V), & \text{if $U > c$}, \\
    0, & \text{otherwise},
  \end{cases}
\end{equation}
where $G(v)$ is function that is non-decreasing in $v$ and $c$ is an
appropriate default threshold that ensures $\mathrm{P}[U > c] = p$, if $p$
denotes the probability of default. $U$ and $V$ may be interpreted as negative
changes in the obligor's assets value at different moments in time, for
instance at default and some months after default respectively.

By
\eqref{eq:product}, with $U = V= \sqrt{\rho}\,X + \sqrt{1-\rho}\,\xi$ the
single risk factor model turns out to be a special case of the potential loss
model.
Note that $F_D^\ast\bigl(\frac{\Phi(\sqrt{\rho}\,X + \sqrt{1-\rho}\,\xi)
-1+p}{p}\bigr)=0$ on the non-default event $\sqrt{\rho}\,X +
\sqrt{1-\rho}\,\xi \le \Phi^{-1}(1-p)$. Hence, in the single risk factor model
the amount of loss equals zero if the the obligor survives. This is in
contrast to the potential loss model, where $V$ and $U$ may be dependent but
need not be identical. As a consequence, $G(V) > 0$ may happen even if $U \le
c$, i.e.\ in case of the obligor's survival. Because of this observation,
\cite{Pykhtin03} called $G(V)$ the \emph{potential loss}.

\cite{Pykhtin03} \citep[cf.\ also][]{Frye00} specifies the vector $(U, V)$ as bivariate
normal vector via
\begin{equation}
\begin{split}\label{eq:U}
  U & = \sqrt{\rho}\,X + \sqrt{1-\rho}\,\xi \qquad\text{and}\\
  V & = \sqrt{\omega}\,X + \sqrt{1-\omega}\,\eta,
\end{split}
\end{equation}
where $X$, $\xi$, and $\eta$ are independent and standard normally distributed
and $\rho, \omega \in (0,1)$ are \emph{asset correlations} (cf.\
Section~\ref{sec:copula}). The constant $c$ in \eqref{eq:A_L} is then given by
$c = \Phi^{-1}(1-p)$.

By \eqref{eq:U} and \eqref{eq:product}, the main difference between the single
risk factor and the potential loss model can be explained as follows.
\eqref{eq:product} shows that in the single loss model the amount of loss and
the default indicator function are \emph{co-monotonous}, i.e.\ they are
connected through the strongest form of stochastic dependence. In contrast, by
\eqref{eq:U} the dependence of the loss amount $G(V)$ and the default
indicator is much weaker. In particular, high values of $G(V)$ may occur even
if the indicator takes the value $0$, i.e.\ there is no default.

\cite{Pykhtin03} suggests to choose $G$ as
\begin{equation}\label{eq:Pykhtin}
  G(v) \ =\ \max\bigl( 1- \exp(-\, \mu - \sigma\,v),\, 0\bigr)
\end{equation}
with $\mu\in\mathbb{R}$ and $\sigma \in (0,\infty)$ fixed.
Together with \eqref{eq:A_L} and \eqref{eq:U}, \eqref{eq:Pykhtin}
completely specifies the potential loss
model. The loss distribution function $F_L$ in this case reads
\begin{equation}\label{eq:FL}
F_L(t) \ = \ \begin{cases}
0, & t < 0,\\
1-p + \Phi\bigl(-\frac{\mu+\log(1-t)}\sigma\bigr) -
\Phi_2\bigl(c,-\frac{\mu+\log(1-t)}\sigma; \sqrt{\rho\,\omega}\bigr), & 0 \le
t
< 1,\\
1, & t \ge 1,
  \end{cases}
\end{equation}
where $\Phi_2(\,\cdot\,,\,\cdot\,;\tau)$ denotes the standard bivariate normal
distribution function with correlation~$\tau$. Observe from \eqref{eq:FL}
that $\mathrm{P}[L = 0] > 1-p$ holds in this model.

At first glance, the potential loss model as specified by \eqref{eq:A_L},
\eqref{eq:U}, and \eqref{eq:Pykhtin} appears to be quite
simple. However, the realized losses are not observed in a form according to
\eqref{eq:A_L} and \eqref{eq:U} but rather via the loss distribution
\eqref{eq:FL}. This loss distribution contains parameters that have to be
estimated in a way that makes estimation a non-trivial task.

The single risk factor model, in contrast to the potential loss model, starts
with the observable loss distribution \eqref{eq:dist}. Complexity enters this
model only in a second step when the conditional expectation
$\mathrm{E}[L\,|\,X = x]$ has to be computed (see \eqref{eq:cond}).  This
calculation of $\mathrm{E}[L\,|\,X = x]$, in turn, is easy in case of the
potential loss model as given by \eqref{eq:A_L} and \eqref{eq:U} since
conditional on $X$ independence of the involved factors holds. With this
observation in mind, we obtain
\begin{align}
  \mathrm{E}[L\,|\,X = x] & =
  \mathrm{P}\bigl[\sqrt{\rho}\,X+\sqrt{1-\rho}\,\xi>\Phi^{-1}(1-p)\,|X=x\bigr]\,
  \mathrm{E}\bigl[G(\sqrt{\omega}\,X+\sqrt{1-\omega}\,\eta)\,|X=x\bigr]
   \notag\\
   & = \Phi\left(\frac{\sqrt{\rho}\,x+\Phi^{-1}(p)}{\sqrt{1-\rho}}\right)
   \int_{-\infty}^\infty G(\sqrt{\omega}\,x+ \sqrt{1-\omega}\,z)\,\varphi(z)\,
   d z. \label{eq:condind}
\end{align}
Given the specification of $G$ by \eqref{eq:Pykhtin}, the integral in \eqref{eq:condind} can be explicitly evaluated
as
\begin{multline}\label{eq:exp_Frye}
\mathrm{E}\bigl[G(\sqrt{\omega}\,X+\sqrt{1-\omega}\,\eta)\,|X=x\bigr]  =
1-
\Phi\left(-\frac{\mu+\sigma\,\sqrt{\omega}\,x}{\sigma\,\sqrt{1-\omega}}\right)\\
- \exp\bigl(-\,\mu - \sigma\,\sqrt{\omega}\,x  + \sigma^2\,(1-\omega)/2\bigr)
\left(1-\Phi\left(-\frac{\mu+\sigma\,\sqrt{\omega}\,x}{\sigma\,\sqrt{1-\omega}}+
\sigma\,\sqrt{1-\omega}\right)\right).
\end{multline}

\end{document}